# Star-forming filament models

**Philip C. Myers**

Harvard-Smithsonian Center for Astrophysics, 60 Garden Street,

Cambridge MA 02138 USA

pmyers@cfa.harvard.edu

**Abstract.** New models of star-forming filamentary clouds are presented, to quantify their properties and to predict their evolution. These 2D axisymmetric models describe filaments having no core, one low-mass core, and one cluster-forming core. They are based on Plummer-like cylinders and spheroids, bounded by a constant-density surface of finite extent. In contrast to 1D Plummer-like models, they have specific values of length and mass, they approximate observed column density maps, and their distributions of column density ($N$-pdfs) are pole-free. Each model can estimate the star-forming potential of a core-filament system, by identifying the zone of gas dense enough to form low-mass stars, and by counting the number of enclosed thermal Jeans masses. This analysis suggests that the Musca Center filament may be near the start of its star-forming life, with enough dense gas to make its first ~3 protostars, while the Coronet filament is near the midpoint of its star formation, with enough dense gas to add ~ 8 protostars to its ~20 known stars. In contrast L43 appears near the end of its star-forming life, since it lacks enough dense gas to add any new protostars to the 2 YSOs already known.

# 1. Introduction

## 1.1. Filamentary clouds

Interstellar clouds are elongated and "filamentary" over a wide range of scales of size and column density. In molecular clouds, such filamentary structure is believed to play an important role in the formation of dense cores and protostars (Molinari et al. 2010, André et al. 2014).

Filamentary clouds are observed by dust extinction of background starlight at optical and near infrared wavelengths, by dust emission at far infrared and submillimeter wavelengths, and by emission in spectral lines tracing a range of gas density. Much of our recent knowledge of filamentary clouds and their properties is based on observations with the *Herschel Space Observatory* of nearby star-forming clouds within a few hundred pc and of more distant infrared dark clouds within a few kpc (Arzoumanian et al. 2011, hereafter A11; Arzoumanian **et al.** 2016, hereafter A16; Malinen et al. 2012; Peretto et al. 2012, Palmeirim et al. 2013, Polychroni et al. 2013, Alves de Oliveira et al. 2014, Koch & Rosolowsky 2015).

Spectral line observations indicate that some filaments which appear monolithic in *Herschel* images can be better understood as "bundles" of closely spaced "fibers" which are distinguished by their velocities and by their incidence of dense cores (Hacar et al 2013, Tafalla & Hacar 2015). However similar observations of other filaments, including the Musca filament, do not indicate multiple fibers (Kainulainen et al. 2015, Hacar et al. 2016).

Simulations of turbulent fragmentation produce a spectrum of filamentary structures, under a variety of initial conditions (Vazquez-Semadeni 1994, Klessen & Burkert 2001, Banerjee et al. 2006, Girichidis et al. 2012, Federrath 2013). Filaments in simulations resemble observed filaments in some ways (Smith et al. 2014, Kirk et al. 2015). There is general consensus that filaments can form from converging flows in regions of supersonic turbulence, and that self-gravitating filament gas can condense into star-forming cores. However it remains unclear how filaments are formed and dispersed, how they supply mass to dense cores, and how their properties affect the star formation rate and protostar mass distribution.

### 1.2. Filament models

The model of filament structure most often compared to observations is the infinite self-gravitating isothermal cylinder (Stodolkiewicz 1963, Ostriker 1964). Non-isothermal infinite cylinders with radially increasing temperature have been considered by Recchi et al. (2013), and infinite polytropic cylinders have been analyzed in both the non-magnetic case (Toci & Galli 2015a) and in the magnetized case (Fiege & Pudritz 2000, Toci & Galli 2015b). Other studies of filament structure, including oscillating filaments, are summarized by Gritschneder et al. (2016).

Models of filament formation and evolution include instability in self-gravitating and magnetized layers (Miyama et al. 1987, Nagai et al. 1998, Hanawa & Tomisaka 2015), formation from converging flow shocks (Pudritz & Kevlahan 2013), and gravitational infall onto filaments (Heitsch 2013). Models of filament fragmentation and core formation are discussed by Larson (1985), Inutsuka & Miyama (1997), and Chen & Ostriker (2015).

In contrast to the above physical models, "descriptive" models of filament structure quantify observed features independent of its dynamical status. A well-known descriptive model is the "Plummer-like" profile of the form

$$n = n_0[1 + (r/r_0)^2]^{-p/2} \qquad (1)$$

where $n_0$ is the maximum density and $r_0$ is a fixed scale length (Plummer 1911). In the limit $r \ll r_0$, $n$ has the constant value $n_0$. When $r \gg r_0$, $n$ declines as a power law in the radial direction, $n = n_0 r^{-p}$, with exponent $p$ indicating the steepness of the density decline.

Plummer-like profiles have been used to describe the density distribution in stellar clusters, globular clusters, starless dense cores, and most recently in filaments (Plummer 1911, King 1962, Nutter et al. 2008, A11, Fischera 2014a, hereafter F14). For $p = 4$ the Plummer-like profile matches the infinite isothermal cylinder (Ostriker 1964), and for $1 < p < 2$ it corresponds to a subisothermal polytropic cylinder (Toci & Galli 2015a). Plummer-like cylinder models are henceforth called "PC models."

PC models fit JCMT/SCUBA submillimeter observations of filament radial column density profiles in Taurus (Nutter et al. 2008), *Herschel* observations in Aquila (A11), and *Herschel* observations in many other nearby regions (A16). The results have been used to estimate filament stability against radial collapse, by applying the stability properties of the infinite isothermal filament, with $p = 4$, to the typical finite filament, with $p \approx 2$. The nearly constant width of observed filaments, combined with their critical line mass for stability, leads to agreement with the column density "threshold" $\sim 7\ 10^{21}\ \mathrm{cm}^{-2}$, above which the gas in nearby molecular clouds tends to harbor young stars (André et al. 2010, Lada et al. 2010, André et al. 2014).

Despite the physical insight they provide, one-dimensional (1D) PC models are too idealized to investigate the finite length and mass of filaments, and their development of dense cores and protostars. PC models are also too idealized to match the distribution of column densities ($N$-pdf) observed in filamentary regions. The constant central column density of a PC implies a pole in its column density distribution ($N$-pdf; F14). In contrast, $N$-pdfs of observed filamentary regions generally have a declining power-law at high column density, with no pole (Schneider et al. 2013, 2015; Kainulainen et al. 2009, 2015).

To better understand how filaments evolve toward star formation, this paper presents three new 2D axisymmetric models of filamentary structure. These models have finite spatial extent and mass, axial structure resembling either spindles or cores, and their $N$-pdfs tend to pole-free power laws at high density. They retain approximately the same mean radial structure as the PC model. They are used to model three observed filamentary clouds, and to estimate their star-forming potential.

In the following sections, Section 2 gives for each model the structure of volume density $n$ and column density $N$, contour maps of $N$, radial $N$-profiles, and the $N$-pdf distribution. Section 3 defines the "star forming zone" (SFZ) of each model, whose gas is dense enough and extended enough to form low-mass stars. It describes a Jeans-like fragmentation model which gives the mean spacing and star formation efficiency of the new stars which the SFZ can produce. Section 4 applies these models to three observed filamentary clouds. It estimates the number of low-mass stars they can form, and compares these numbers to their already known population of young stars. Section 5 summarizes the paper and discusses limitations and applications.

**2.1. Axisymmetric 2D filament models**

The models presented here are axisymmetric, where the volume density depends on the radial extent $r$ from the symmetry axis and on the axial distance $|z|$ from the center. Each of the models has density depending on radius as $r^{-2}$, in the limit where $r$ is much greater than the scale length $r_0$, corresponding to the $p = 2$ case for a Plummer cylinder.

More complex structures, including magnetized filaments (Fiege & Pudritz 2000), bundles of fibers (Hacar et al. 2011), and filament networks (Busquet et al. 2013) are beyond the scope of this paper. The models presented here are condensed in the radial and the axial directions. They have no simple equilibrium interpretation, in contrast to simple PC models, which are radially condensed but axially uniform. The dynamical evolution of these nonequilibrium models may be a useful application for numerical simulations (e.g. Nelson & Papaloizou 1993, Sigalotti & Klapp 2001, Burkert & Hartmann 2004).

Sections 2.3.1 - 2.3.2 give expressions for density and column density for each model considered. These expressions are used to generate the column density maps, radial profiles, and $N$-pdfs used in later sections. Readers more interested in results may prefer to skip to Section 2.4.

**2.2. Cylindrical models**

This section gives expressions for $n$ and $N$ for models whose normalized radial density structure is cylindrical, i.e. $n(r, z)/n(0, z)$ is independent of axial position $z$. These consist of the 1D Plummer Cylinder (PC; Nutter et al. 2008, A11), the 1D Truncated Plummer Cylinder (TPC; F14), and the 2D Truncated Plummer-Plummer Cylinder (TPPC) introduced in this work.

2.2.1. Plummer Cylinder (PC) and Truncated Plummer Cylinder (TPC). The density structure of the $p$ = 2 PC is

$$n_{PC} = n_0\left[1+\left(r/r_0\right)^2\right]^{-1} \tag{2}$$

for scale length $r_0$, central density $n_0$, and cylindrical radius $r = (x^2+y^2)^{1/2}$ in the range $0 < r < R$, where $R \rightarrow \infty$. The corresponding column density is obtained by integrating equation (2) along the $y$ - axis, assuming that the symmetry axis coincides with the $z$ - axis:

$$N_{PC} = N_{bk} + \pi n_0 r_0\left[1+\xi^2\right]^{-1/2} \tag{3}$$

where $\xi \equiv x/r_0$ and where a constant background column density $N_{bk}$ is assumed. Here $N_{PC}$ declines from its maximum $N_{PC} = N_{bk} + \pi n_0 r_0$ at $x = 0$ to its minimum $N_{PC} = N_{bk}$ at $x = \infty$.

The TPC has the same density structure as the PC in equation (2),

$$n_{TPC} = n_0\left[1+\left(r/r_0\right)^2\right]^{-1} \tag{4}$$

but its maximum radius $R$ is finite rather than infinite. Then equation (4) defines the bounding surface in terms of the minimum density $n_{\min}$ by

$$\Xi = \left[\frac{n_0}{n_{\min}} - 1\right]^{1/2} \qquad (5)$$

where $\Xi \equiv R/r_0$ is the maximum value of $\xi = x/r_0$ at the boundary, where $x = R$ and $y = 0$.

The TPC column density is obtained by integrating its volume density along the $y$-axis, giving

$$N_{TPC} = N_{bk} + \frac{2n_0 r_0}{\left(1+\xi^2\right)^{1/2}} \tan^{-1}\left\{\left[\frac{\Xi^2 - \xi^2}{1+\xi^2}\right]^{1/2}\right\} \qquad (6)$$

Here $N_{TPC}$ declines from its maximum $N_{TPC} = N_{bk} + 2n_0 r_0 \tan^{-1}\Xi$ at $x = 0$, to its minimum $N_{TPC} = N_{bk}$ at $x = R$. Note that equation (6) reduces to equation (3) when $R >> r_0$, as expected.

2.2.2. Observational constraints on 2D models. To match observed filament properties, the 2D models presented here have density functions $n(r, z)$ subject to three observational constraints. (1) The $N$-pdf should be pole-free as in column density observations of filamentary regions (Schneider et al. 2015, 2016). This condition is met when the axis density profile $n(0, z)$ has sufficient variation with axial position $z$. (2) The mean radial column density profile should be well-fit by a PC model with $p \approx 2$, as is typical of filaments observed in nearby clouds with *Herschel* (A16). This condition is met when the density $n$ varies approximately as $[1 + (r/r_0)^2]^{-1}$. (3) The contour maps of column density $N$ should have high-$N$ contours of approximately ellipsoidal shape, elongated along the long axis, to resemble observed contour shapes of cores and filamentary ridges. This condition is met when the axial profile $n(0, z)$ depends on $z$ with a form similar to the TPC dependence of $n(r, 0)$ on $r$.

2.2.3. TPPC cylinder model. The above constraints on 2D cylinder models are met in a "TPPC" cylinder model

$$n_{TPPC} = \frac{n_0}{\left[1 + \xi^2 + \eta^2\right]\left[1 + \left(\zeta/a\right)^2\right]} \quad (7)$$

where $n_0$ is the peak density at the origin, and where $\eta \equiv y/r_0$ and $\zeta \equiv z/r_0$. Here $a$ is the aspect ratio of the bounding surface, which is obtained from equation (7) by setting

$$\left\{\left[1 + \xi^2 + \eta^2\right]\left[1 + \left(\zeta/a\right)^2\right]\right\}_b = 1 + \Xi^2 \quad (8)$$

where

$$\Xi = \left[\frac{n_0}{n_{\min}} - 1\right]^{1/2} . \quad (9)$$

Here $\Xi$ is the maximum value of $\xi$, when $\eta = \zeta = 0$, or equivalently $\Xi$ is the maximum value of $\eta$, when $\xi = \zeta = 0$. Then equation (8) gives the maximum value of $\zeta$, denoted $Z$, as

$$Z = a\Xi \tag{10}$$

whence the bounding surface aspect ratio is $Z / \Xi = a$. Note that the bounding surface is not a simple ellipsoid; instead it resembles a prolate figure with a central bulge.

The column density of the TPPC model is obtained by integrating equation (7) along the $y$-axis, within the limits given in equation (8), giving

$$N_{TPPC} = N_{bk} + \frac{2n_0 r_0}{\left(1+\xi^2\right)^{1/2}\left[1+\left(\zeta/a\right)^2\right]} \tan^{-1}\left\{\left[\frac{\left(1+\Xi^2\right)/\left(1+\xi^2\right)}{1+\left(\zeta/a\right)^2} - 1\right]^{1/2}\right\}. \tag{11}$$

Equation (11) indicates that the column density declines from its maximum $N_{TPPC} = N_{bk} + 2n_0 r_0 \tan^{-1}\Xi$ at $(x, z) = (0, 0)$ to its minimum $N_{TPPC} = N_{bk}$ at the bounding column density contour, defined by

$$\left\{\left[1+\xi^2\right]\left[1+\left(\zeta/a\right)^2\right]\right\}_b = 1 + \Xi^2 \quad . \tag{12}$$

Comparison of equations (8) and (12) indicates that the bounding column density contour is simply the intersection of the bounding surface and the plane of the sky (the $x = 0$ plane). Thus the aspect ratio of the bounding contour is equal to $a$ as in equation (10).

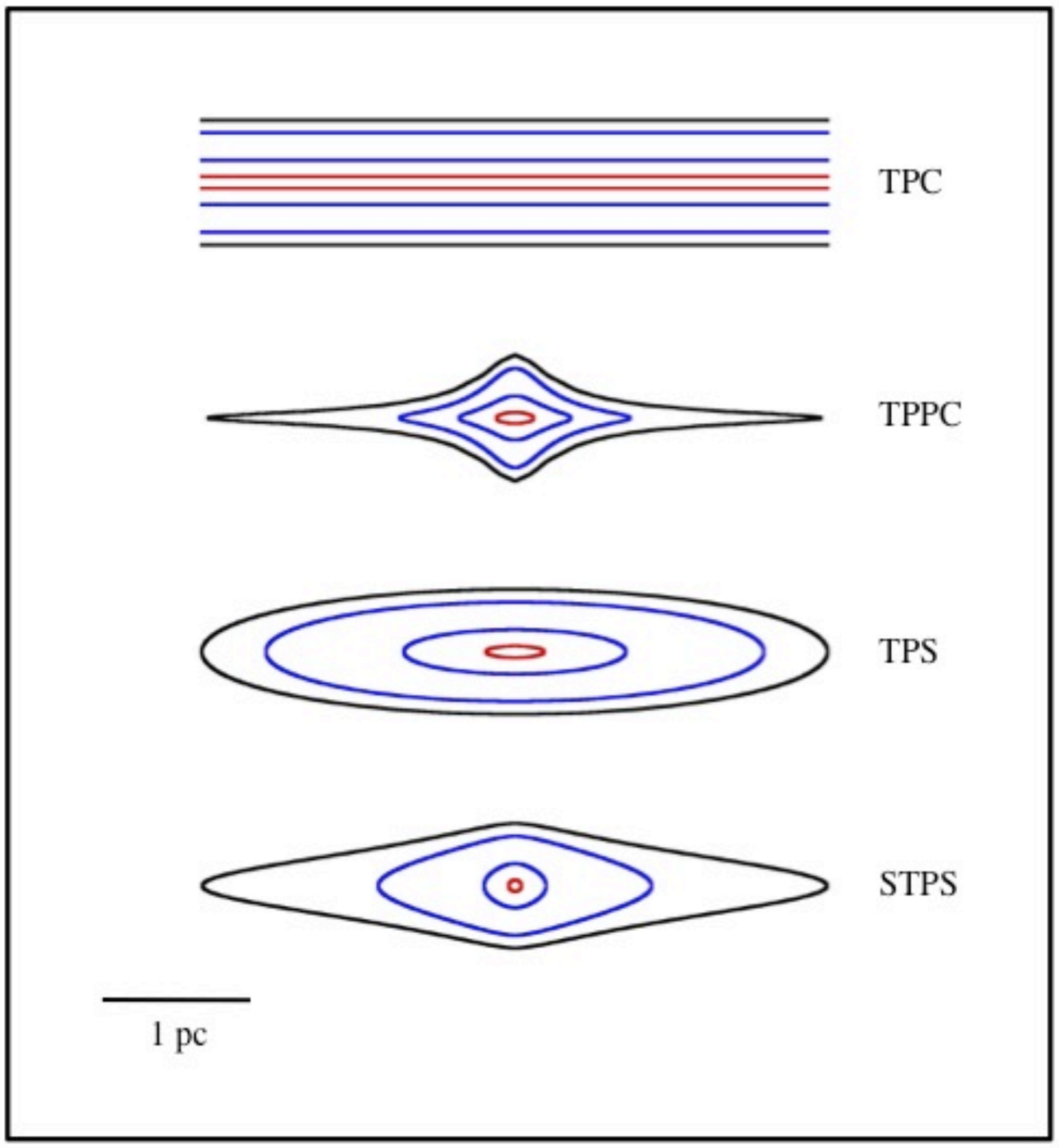

**Figure 1.** Contour maps of column density for the truncated filament models TPC (truncated Plummer cylinder), TPPC (truncated Plummer-Plummer cylinder), TPS (truncated prolate spheroid), and STPS (stretched truncated prolate spheroid). Each model assumes parameter values $r_0$ = 0.04 pc, $n_0$ = $10^5$ $cm^{-3}$, $R$ = 0.4 pc, and maximum aspect ratio $a$ = 5. Contours of constant column density are drawn at 1, 3, 9, and 27 $10^{21}$ $cm^{-2}$.

To aid in visualizing the models, Figure 1 shows contour maps of column density for the TPC, TPPC, TPS, and STPS models. They are all based on the same scale length $r_0$ = 0.04 pc, peak density $n_0$ = $10^5$ $cm^{-3}$, maximum radius in the radial direction $R$ = 0.4 pc, and maximum radius in the axial direction 2 pc. These values are chosen to be typical of core and filament properties in nearby star-forming regions.

In the TPPC model, the column density contours change shape, from small radii to large radii. The contour aspect ratio increases with increasing radius, and in each quadrant the contour shape is concave at small radius and convex at large radius. The bounding contour shape approaches that of an ellipse only in the limit as $\zeta$ approaches its maximum value $Z$. The aspect ratio of this limiting ellipse is $a\Xi$; it is more elongated than the aspect ratio $a$ of the TPPC bounding contour, and more elongated than the aspect ratio $a/\sqrt{2}$ of the innermost contour. These properties make the TPPC model a useful description of an elongated filament with a large central bulge, such as the integral-shaped filament in Orion A North (e.g. Salji et al. 2015).

**2.3 Spheroidal models** This section describes Truncated Prolate Spheroid (TPS) models, whose column density contours are ellipses of constant aspect ratio, and Stretched Prolate Spheroid (STPS) models, whose contours are approximately ellipses whose aspect ratio increases with radius.

Such spheroidal models with constant aspect ratio may be useful to describe filaments whose high-column-density gas is mostly concentrated in elongated ridges, as in Chamaeleon I (De Oliveira et al. 2014) or Musca (Kainulainen et al. 2015). Spheroidal models with outwardly increasing aspect ratio may also describe a filament which harbors a single low-mass core, as in L43 (Mathieu et al. 1988, Chen et al. 2009). In spheroidal models, the density depends on axial and radial coordinates which are summed in quadrature. They are not separable as in the above TPPC cylinder model. Therefore their outer column density contours are ellipses in contrast to the TPPC model.

Oblate and prolate spheroidal models have been widely used as analytic descriptions of galaxy structure (Binney & Tremaine 1987). In some observed cases the ellipticity is not constant with radius (King 1978), requiring oblate models which are not strictly spheroidal (Bohn 1983).

Prolate spheroidal models have been used to fit column density structures in molecular cloud images, to obtain the volume density pdf of the cloud (Kainulainen et al. 2014). Similar models have been used as initial states for numerical calculations of collapse and fragmentation (Nelson & Papaloizou 1993, Sigalotti & Klapp 2001). Prolate spheroidal models have been studied to a lesser degree than cylindrical models. Prolate spheroidal models of varying ellipticity, presented here, do not appear to have been studied previously.

2.3.1. Truncated Prolate Spheroid (TPS)

The spheroidal models used here are Plummer-like, in order to approximate the radial profiles derived from *Herschel* observations (A11, A16). Their volume density $n$ depends on space coordinates as

$$n_{TPS} = \frac{n_0}{1 + \xi^2 + \eta^2 + \left(\zeta/a\right)^2} \tag{13}$$

where as in Section 2.2 the fixed parameters are the peak volume density $n_0$ and the radial scale length $r_0$. The constant parameter $a$ is the ratio of the maximum radii in the axial and radial directions. It is assumed that $a > 1$, so that the spheroid is prolate.

The density is truncated at a minimum value $n_{\min}$ by a constant-pressure medium. Then equation (13) sets the bounding surface, which is a prolate spheroid satisfying

$$\left[\xi^2 + \eta^2 + (\zeta/a)^2\right]_b = \Xi^2 \tag{14}$$

where $\Xi$ is the maximum value of $\xi$, as defined in equation (5).

The column density in the $y$-direction is obtained by integrating the density in equation (13), giving

$$N_{TPS} = N_{bk} + \frac{2n_0 r_0}{\sqrt{1+\mu^2}} \tan^{-1}\sqrt{\frac{\Xi^2 - \mu^2}{1+\mu^2}} \tag{15}$$

where the dimensionless coordinate $\mu$ is the quadrature sum of the normalized coordinates in the $x$ and $z$ directions,

$$\mu \equiv \left[\xi^2 + (\zeta/a)^2\right]^{1/2} \tag{16}$$

and where $0 \le \mu \le \Xi$. The column density declines from its maximum $N_{TPS} = N_{bk} + 2n_0 r_0 \tan^{-1}\Xi$ at $(x, z) = (0, 0)$ to its minimum $N_{TPS} = N_{bk}$ at the bounding column density contour, defined by

$$\left[\xi^2 + (\zeta/a)^2\right]_b = \Xi^2 \quad . \tag{17}$$

Equations (15) - (17) show that $N_{TPS}$ depends on $\xi$ and $\zeta$ only through $\mu$. The contours of column density are concentric ellipses of aspect ratio $a$, nested in the bounding contour, as shown in Figure 1. As $a \rightarrow 1$ the column density of the TPS reduces to that for the truncated Plummer sphere (F14, equation A.11).

2.3.2. Stretched Truncated Prolate Spheroid (STPS). Many observed filaments have at least one embedded low-mass core, a significant local maximum of column density whose radial width is similar to that of its host filament, and whose aspect ratio is less than ~2. For these systems the PC, TPC, and TPS models are not useful because they lack a significant core, while the TPPC model is not useful because its core is too extended.

A formulation which meets the three constraints of Section 2.2.2 has the aspect ratio $a$ in equation (10) increasing with $z$, so that column density contours progress from nearly round near the center where $z = 0$, to elongated near the extreme values $\pm z_{max}$. This result can be achieved if $a$ increases linearly from $a_{min}$ when $\zeta = 0$ to $a_{max}$ when $|\zeta| = \zeta_{max}$, i.e.

$$a(\zeta) \equiv a_{min} + (a_{max} - a_{min})|\zeta| / \zeta_{max} \qquad (18)$$

With this variable aspect ratio, the expression for the STPS density becomes

$$n_{STPS} = \frac{n_0}{1 + \xi^2 + \eta^2 + \left[\zeta / a(\zeta)\right]^2} \quad , \qquad (19)$$

where the bounding surface satisfies

$$\left\{\xi^2 + \eta^2 + \left[\zeta / a(\zeta)\right]^2\right\}_b = \Xi^2 \; . \qquad (20)$$

The column density becomes

$$N_{STPS} = N_{bk} + \frac{2n_0 r_0}{\sqrt{1+\nu^2}} \tan^{-1} \sqrt{\frac{\Xi^2 - \nu^2}{1+\nu^2}} \qquad (21)$$

where the dimensionless coordinate $\nu$ for the stretched spheroid is

$$\nu \equiv \left[\xi^2 + \left[\zeta / a(\zeta)\right]^2\right]^{1/2} \quad , \qquad (22)$$

and where the bounding column density contour satisfies

$$\left\{\xi^2 + \left[\zeta / a(\zeta)\right]^2\right\}_b = \Xi^2 \quad . \qquad (23)$$

As in the TPC, TPS, and TPPS models, the column density declines from its maximum value $N_{STPS} = N_{bk} + 2n_0 r_0 \tan^{-1} \Xi$ at $x =$ z = 0 to its minimum $N_{STPS} = N_{bk}$ at the bounding contour.

The STPS contours in Figure 1 resemble ellipses which have been "stretched" along the long axis by a factor which increases with distance from the center. Thus this model is denoted STPS. Equations (18) and (23) indicate that the aspect ratio of the bounding contour equals $a_{\max}$. The aspect ratio $a_{\mathrm{HM}}$ of the half-maximum contour for the STPS is much closer to unity than for the TPS, as equations (10) and (23) show. The STPS model may give a useful description of a low-mass core with nearly round contours, embedded in a more elongated filament.

### 2.4. Radial *N*-profiles

The mean radial column density profile $\overline{N}(x)$ of a filamentary cloud has been used to characterize the radial structure of observed clouds, to compare from cloud to cloud and to compare with theoretical models and simulations (A11, A16, Koch & Rosolowsky 2015, Kirk et al. 2015, Smith et al. 2014). This section compares $\overline{N}(x)$ for the TPC, TPPC, TPS, and STPS models. It finds that the mean profile shapes are highly similar for all the models, for relative amplitudes greater than ~20% of the peak amplitude. This similarity of shape extends over a greater range for the TPC, TPPC, and TPS models, down to $\sim$ 10% of the peak amplitude.

The typical analysis procedure first defines the "spine" of the filament taking into account its departures from a straight line, using one of several available image processing algorithms (Koch & Rosolowsky 2015). Then one obtains the mean column density $\overline{N}(x) = \int dz N(x,z) / \int dz$, where $x$ is the distance perpendicular to the local spine direction, at each of a sequence of equally spaced values of $z$ along the spine.

The mean radial structure profile $\overline{N}(x)$ is typically fit with a Plummer cylinder model based on equation (1), out to a radial distance where the profile merges with the local background. Usually it is possible to obtain a good fit where the scale length $r_0$ is a few 0.01 pc and where the density exponent $p$ lies in the range 1.5 - 2 (A11, A16). However these parameters are not independent (Kirk et al. 2015, Smith et al. 2014).

The mean radial profiles of the TPC, TPPC, TPS, and STPS models were compared with those of the PC model, by calculating $N(x, z)$ at ten values of $z$ extending in uniform steps of 0.05 pc along the z-axis. The profiles $N(x, z)$ were combined in an unweighted average. At points where $x$ extended beyond the truncation boundary, $N$ was set to zero before averaging with neighboring scans. Each mean profile $\overline{N}(x)$ was normalized to its maximum value, giving the normalized mean column density profile $\nu(x) \equiv \overline{N}(x) / \overline{N}(0)$.

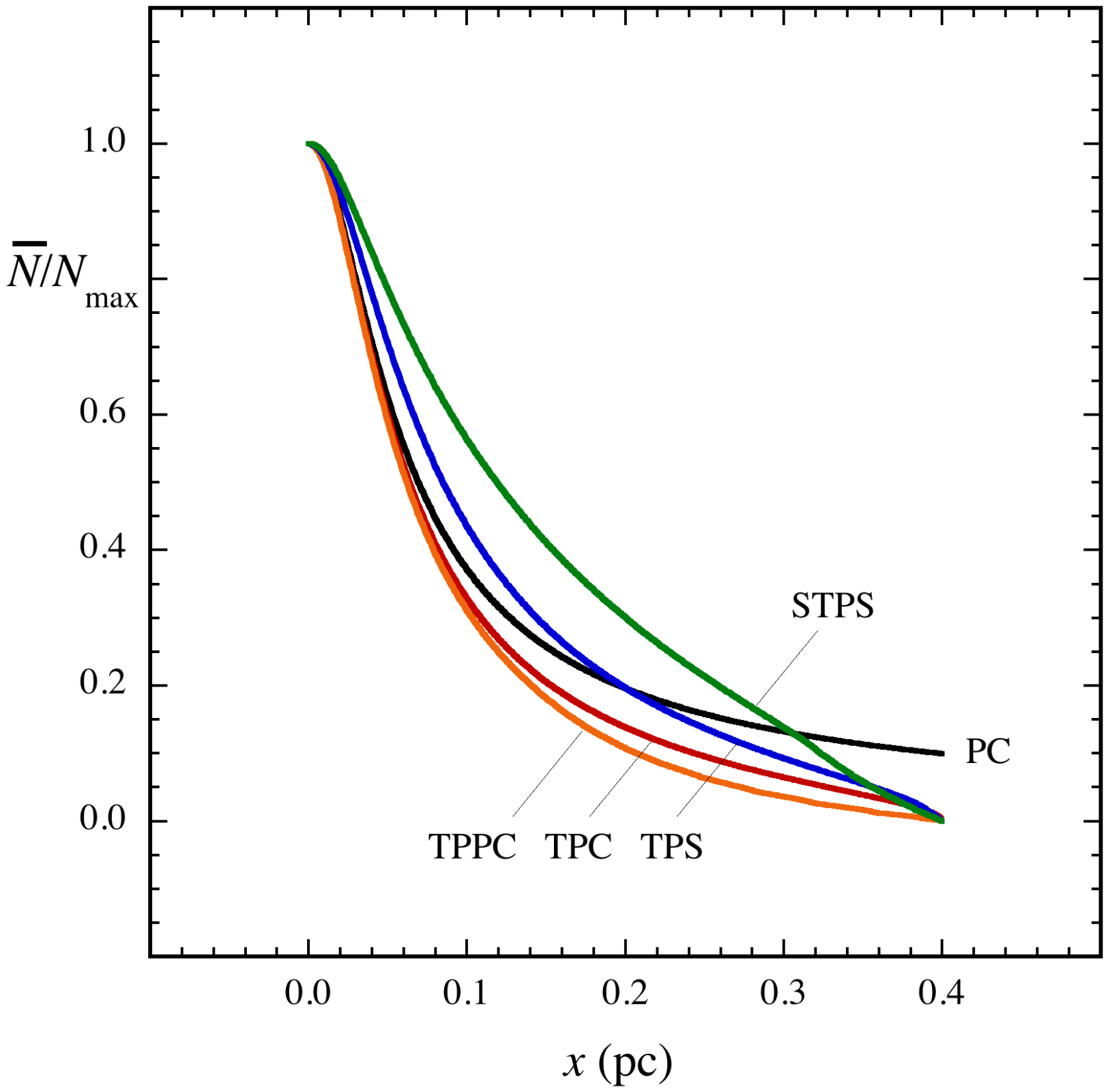

**Figure 2.** Mean radial profiles of column density for the five models PC, TPC, TPPC, TPS, and STPS, normalized to their maximum value, for fixed radial scale length $r_0 = 0.04$ pc.

The width and shape of the normalized profile $\nu(x)$ are compared with those of the PC profile in Figures 2 and 3. Figure 2 shows the five profiles when each model has the same scale length $r_0 = r_0(\mathrm{PC}) = 0.04$ pc, and Figure 3 shows the profiles when the model scale lengths are adjusted so that each profile has the same HWHM width $x_{1/2} = x_{1/2}(\mathrm{PC}) = 0.069$ pc. These figures show that the TPC, TPPC, TPS, and STPS model profiles have similar widths and shapes to those of the PC model, and to each other. This similarity can be expected since each model has the same basic dependence of density on radius as $n \sim [1 + (r/r_0)^2]^{-2}$. Furthermore each model has the identical radial column density profile at $z = 0$, as can be seen from equation (11), (15), and (21), or by inspection of the contours in Figure 1.

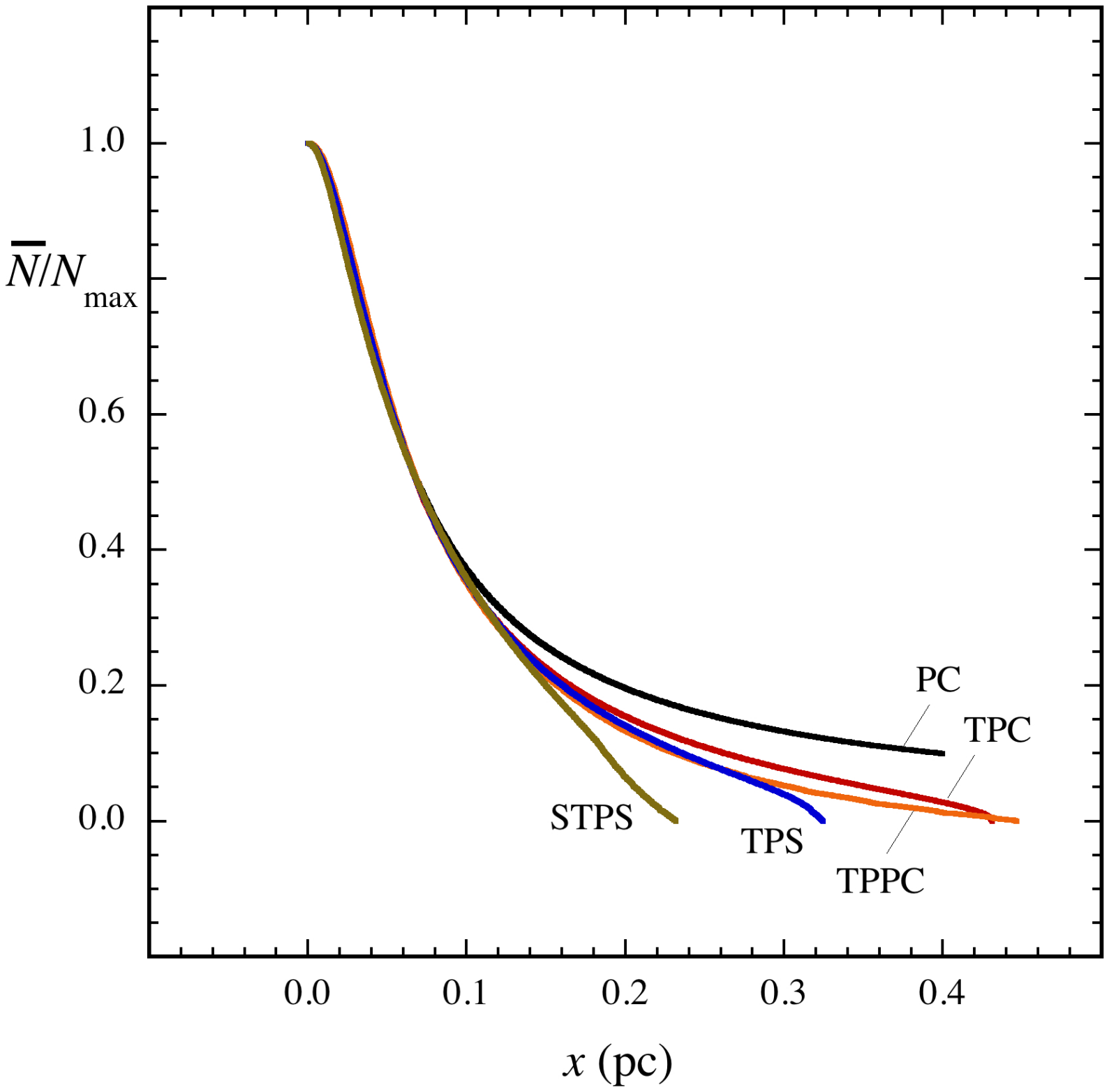

**Figure 3.** Mean radial profiles of column density for the five models PC, TPC, TPPC, TPS, and STPS, normalized to their maximum value, for scale lengths $r_0$ adjusted so that each profile has the same HWHM $x_{1/2}$ = 0.069 pc.

Figure 2 shows that for fixed scale length, the HM radii $x_{1/2}$ of the four truncated models all lie within a factor of 2 of the HM radius of the PC, from $x_{1/2}$(TPPC)= 0.90 $x_{1/2}$(PC) to $x_{1/2}$(STPS) = 1.72 $x_{1/2}$(PC). The average ratio of the TPC, TPPC, TPS, and STPS width to the PC width is 1.2. This variation in width arises mainly because the width of each model profile scales slightly differently with scale length. Thus each model can match the mean width of an observed profile with a simple adjustment of its scale length. Alternatively if the observed profile has a well-defined central maximum it may be possible to match the width of the central profile instead of the mean profile.

Figure 3 shows that model profiles having the same HM width also have the same basic shape as the PC profile, for relative amplitudes above ~ 0.2 or equivalently for radial extent within the first 2-3 HM radii from the filament axis. For larger radial extents the STPS profile diverges most from the PC profile. In contrast the TPS, TPPS, and TPC profiles remain similar to each other, down to lower relative amplitude ~ 0.1. At this level their departure from the PC model is due mainly to the difference between truncation at finite radial distance (TPC, TPPS, TPC) and infinite radial distance (PC).

## 2.5. *N*-pdfs

This section presents *N*-pdfs for the TPC, TPPC, TPS, and STPS models, to compare with the typical observed properties of a well-defined peak and a pole-free power-law decline at high *N*. The main result is that each of the new models matches these properties, in contrast to the TPC model, whose pole at high *N* conflicts with observations.

The number distribution of log column density in a region, or its *N*-pdf, is a diagnostic of the dense gas and star-forming properties in a molecular cloud (Schneider et al. 2013, 2015; Kainulainen et al. 2009, 2015; Federrath et al. 2013). The *N*-pdf is defined as $Np(N)$, where $p(N)$ is the probability density that the column density lies between $N$ and $N + dN$. Many *N*-pdfs are observed to have a well-defined peak column density and a negative power-law slope at high *N*. A shallower slope is associated with a greater degree of star formation in a region (Sadavoy et al. 2014, Stutz & Kainulainen 2015). The value of the slope has been interpreted as an indicator of the dynamical status of its dense gas (Federrath et al. 2013, Girichidis et al. 2014), and of its degree of central concentration (Myers 2015).

Some $N$-pdfs cannot be expected to represent the properties of any simple structure, because their corresponding observed regions harbor too many clouds of diverse structure. However, obervations with finer resolution and sensitivity make it possible to obtain the $N$-pdfs of regions whose emission is dominated by one or a few filaments, such as the IC 5146 region (Schneider 2015, pers. comm.), or the Musca filament (Kainulainen 2016). Thus it is useful to compute the $N$-pdfs of the filamentary models presented here, for comparison with observations.

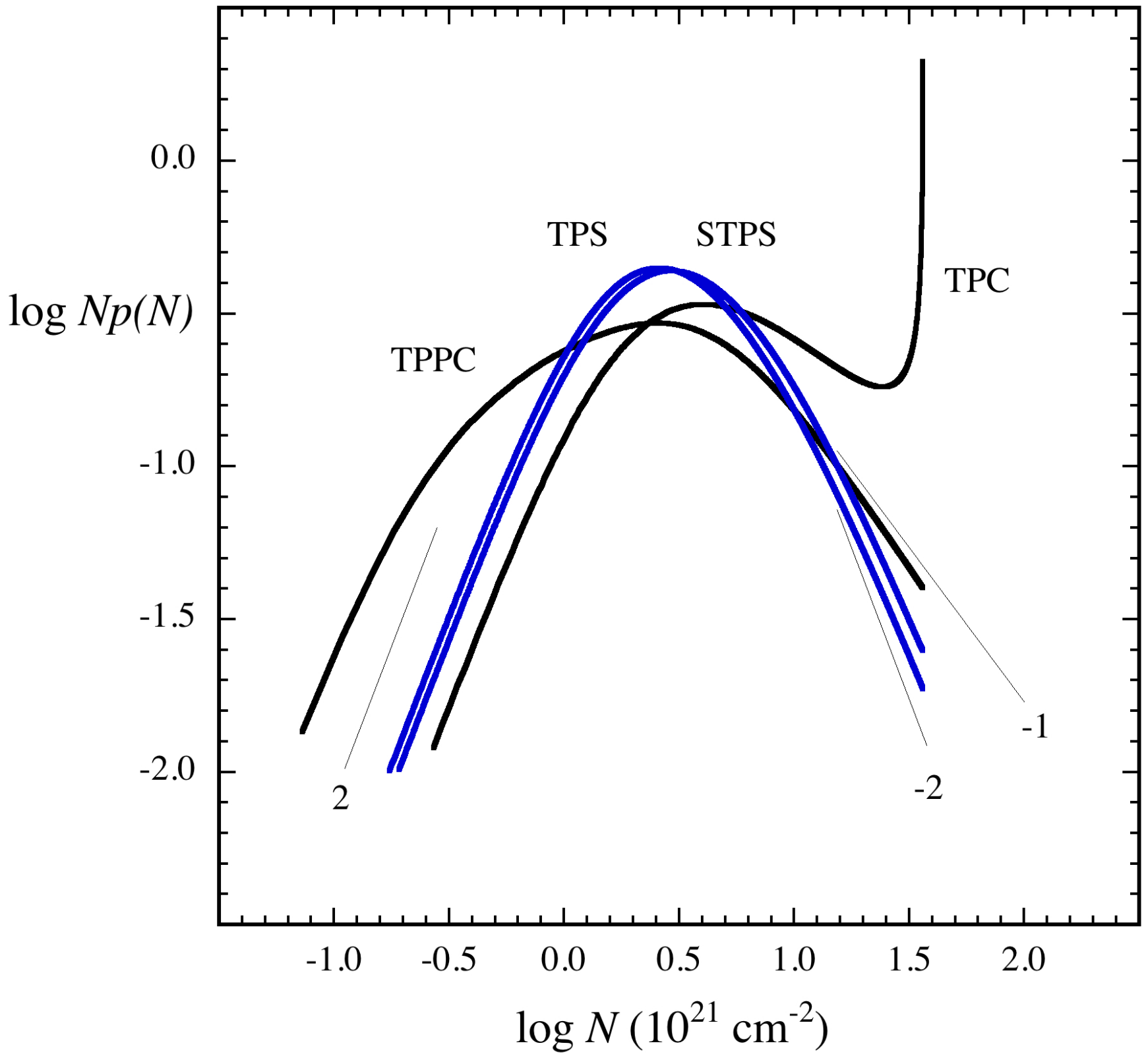

**Figure 4.** Column density distributions ($N$-pdfs) for the truncated models TPC, TPPC, TPS, and STPS. *Black* curves indicate cylindrical models which are axially uniform (TPC) or axially concentrated (TPPC). *Blue* curves indicate spheroidal models having constant aspect ratio (TPS) or aspect ratio increasing outward from the center (STPS). *Faint lines* indicate the asymptotic slopes expected for pure Plummer cylinders (-1) or spheres (-2) at high density, or for either at low density (2).

The method of $N$-pdf calculation is similar to that described in F14 and in Myers (2015). However it differs because the TPPC, TPS, and STPS models have neither purely radial nor purely cylindrical symmetry. Therefore $p(N)$ was found by dividing the total area $A$ within the bounding contour into axial "slices" of area $2x_{max}(z)dz$. Within each slice, the area was found between $z$ and $z + dz$ and between the contours of $N$ and of $N + dN$. This area was integrated numerically over all $z$ to give the differential area between $N$ and $N + dN$. Then the probability $p(N)dN$ that the column density lies between $N$ and $N + dN$ is the ratio of differential to total area, or equivalently,

$$p(N) = -\int_0^{z\,\max(N)} dz\, \frac{dx}{dN}(z) \Big/ \int_0^{z\,\max(N\mathrm{bk})} dz\, x_{\max}(z) \quad . \tag{24}$$

Here $z_{max}(N)$ is the maximum value of $z$ in the contour of constant $N$, $z_{max}(N_{bk})$ is the maximum value of $z$ at the bounding contour, and $x_{max}(z)$ is the value of $x(z)$ at the bounding contour. For each calculation of the $N$-pdf, the probability density $p(N)$ was found to satisfy the normalization condition $\int p(N)dN = 1$.

In equation (24) the derivative $dx/dN$ was obtained analytically from equations (11), (15), and (21) with an approximation useful when $\Xi$ is sufficiently large. In the power series representations for $\tan^{-1}(x)$ and for $\tan^{-1}(x_{max})$ (Gradshteyn & Ryzhik (1980, #1.644 1), one may assume that the ratio of hypergeometric functions $F[1/2, 1/2; 3/2; x^2/(1 + x^2)]$ to $F[1/2, 1/2; 3/2; x_{max}^2/(1 + x_{max}^2)]$ is negligibly different from unity. Then the function

$$f = \frac{1}{\left(1 + \xi^2\right)^{1/2}} \tan^{-1}\left\{\left[\frac{\Xi^2 - \xi^2}{1 + \xi^2}\right]^{1/2}\right\} \Big/ \tan^{-1}(\Xi) \tag{25}$$

can be approximated by

$$f_{\rm app} = \frac{1}{\Xi}\left[\frac{\Xi^2 - \xi^2}{1 + \xi^2}\right]^{1/2} \quad . \qquad (26)$$

where $0 \le \xi \le \Xi$ and $0 \le f \le 1$. Here $f_{\rm app}$ matches $f$ exactly at $\xi = 0$ and at $\xi = \Xi$, and $f_{\rm app}$ overestimates $f$ slightly for all other values of $\xi$. This approximation has uncertainty smaller than the usual observational uncertainties. When $\Xi = 10$, the mean of $f_{\rm app} - f$ is 2% of the range of $f$.

Figure 4 shows *N*-pdfs for the cylindrical TPC and TPPC models, and for the spheroidal TPS and STPS models, computed as described above for the same parameters as in Figure 1. The TPPC, TPS and STPS models have pole-free *N*-pdfs with power-law behavior at high *N*, in contrast to the TPC which has a pole at the central column density. In addition, these new *N*-pdfs have peaks and slopes which vary only slightly from *N*-pdfs of their purely spherical and cylindrical counterparts.

The *N*-pdf for the axially truncated Plummer-Plummer cylinder (TPPC) has no pole, because in this case the peak column density for each slice differs from the peak column densities of all the other slices. The resulting probability *p(N)* contains an infinitesimally small pole at each *N*, because *p(N)* is an average of the pole probability for each slice with smaller probability values from all the other slices. Thus the TPC *N*-pdf has a pole because its on-axis density is uniform, while the TPPC *N*-pdf is pole-free because its on-axis density is sufficiently nonuniform. The TPC and STPC *N*-pdfs are similarly pole-free because of their axial nonuniformity.

All of the *N*-pdfs in Figure 4 have a power law slope of 2 at low *N*, and a local maximum near $N = 3\ 10^{21}\ {\rm cm}^{-2}$, or approximately at $N = 2n_0r_0/\Xi$. Each of these properties is a result known for pure Plummer cylinders and spheres (F14). Evidently the departures of these new models from pure Plummer cylinders and spheres do not significantly change these properties.

At high *N,* the *N*-pdf slopes of the TPC and TPPC models tend toward -1 as expected for an axially uniform Plummer cylinder (PC) with infinite radial extent (F14). At low *N,* the TPC and TPPC *N*-pdf slopes are each 2, but the *N*-pdf of the TPPC extends to lower column density than does the *N*-pdf of the TPC. This difference occurs because the TPPC density declines from its peak value in both the axial and radial directions, while the TPC density declines only in the radial direction.

## 3. Star-forming zone

Section 2 shows that the TPPC, TPS and STPS models can approximate the large-scale density structure of simple filamentary clouds, without violating observational constraints on their *N*-profiles and *N*-pdfs. In turn, this density structure can be used to estimate a cloud's capacity to form new stars. This section describes for each model its "star-forming zone" (SFZ) and gives an estimate of the number of protostars the SFZ can form.

### 3.1. Star-forming zone of a model cloud

A star-forming zone is defined here as a region of a molecular cloud dense enough and extended enough to form stars of typical mass, in contrast to surrounding cloud gas which is less dense and which does not form such stars. These SFZ properties are based on observations of nearby regions of low-mass star formation. On small scales, dense cores harboring protostars have mean density ~ 3 $10^4$ $cm^{-3}$ over ~ 0.05 pc according to $NH_3$ line observations and dust continuum emission (Myers & Benson 1983, Beichman et al. 1986, Enoch et al. 2006, Sadavoy et al. 2010). On larger scales, cloud regions extending up to ~ 1 pc are associated with multiple cores and young stellar objects when their mean column density exceeds ~ 6 $10^{21}$ $cm^{-2}$, based on submillimeter dust emission (André et al. 2010), and on near-infrared extinction of background stars (Lada et al. 2010). These observations indicate a relatively sharp increase in the incidence of young stars above a "threshold" column density.

A model SFZ can match these properties if its smallest possible version has the mass and extent of a single dense core. The core model adopted here is the critically stable isothermal sphere (BE sphere, Bonnor 1956, Ebert 1955) which forms a star having the mean mass of the initial mass function (IMF), $\overline{M}_{IMF} = 0.36\ M_{\rm Sun}$ (Weidner & Kroupa 2006). This particular BE sphere is here called the "Mean-IMF-Sphere" or "MIS." For simplicity MIS properties are written with subscript *S*, and SFZ properties are written with subscript *Z*.

The MIS properties depend on its core-star efficiency $\varepsilon_{\rm S}$ and temperature $T_{\rm S}$. Here $\varepsilon_{\rm S} = M_{\rm star}/M_{\rm S}$ is assumed equal to 0.35, the mean of values obtained from core mass functions in the Pipe Nebula (Alves et al. 2007) and in the Aquila complex (Könyves et al. 2015). The MIS mass is then $M_{\rm S} = 1.0\ M_{\rm Sun}$. Star-forming gas may tend to fragment into cores of about this mass, due to the thermal coupling of gas and dust (Larson 2005), or due to converging magnetized flows (Chen & Ostriker 2015). The temperature $T_{\rm S}$ is close to 10 K in nearby regions of isolated low-mass star formation. In young clusters and in regions of more massive star formation, the temperature of star-forming gas may exceed 20 K (Jijina et al. 1999, Rosolowsky et al. 2008, Foster et al. 2009).

A MIS with $M_{\rm S} = 1.0\ M_{\rm Sun}$ and $T_{\rm S} = 10$ K matches the observed star-forming properties cited above, since its radius, boundary density and mean column density are respectively $R_{\rm S} = 0.05$ pc, $n_{\rm min} = 1.3\ 10^4\ {\rm cm}^{-3}$, and $\overline{N}_{\rm S} = 6.8\ 10^{21}\ {\rm cm}^{-2}$. Thus the SFZ model adopted here is a centrally concentrated region dense enough and extended enough to form at least one star of typical IMF mass, whose radius in each direction is at least $R_{\rm S}$, whose minimum bounding density is $n_{\rm min}$, and whose mean column density is at least $\overline{N}_{\rm S}$. It may have any closed shape and its extent may be much greater than $R_{\rm S}$.

For fixed mass and increasing temperature, the MIS becomes smaller and denser. If $T$ = 20 K, the MIS properties are $R_S$ = 0.025 pc, $n_{min}$ = 1.0 $10^5$ cm$^{-3}$, and $\overline{N}_S$ = 27 $10^{21}$ cm$^{-2}$. The SFZ mass $M_Z$ and volume $V_Z$ are obtained by integrating the model cloud density over the SFZ. The mean mass density of the SFZ is defined as $\overline{\rho}_Z = M_Z / V_Z$. The SFZ "concentration" is the ratio of its mean to minimum density, denoted $q_Z \equiv \overline{\rho}_Z / \rho_{min}$.

3.2 Fragmentation of the star-forming zone

To estimate how many stars of typical mass the SFZ can form, it is necessary to specify its fragmentation. A simple model adopted here is based only on properties of the SFZ and MIS discussed above, and on mass and volume conservation. It assumes (1) the SFZ mass and volume are the same before and after fragmentation, (2) the SFZ fragments only into MISs and uniform gas of density $n_{min}$, and (3) each MIS collapses into a star of final mass $\overline{M}_{IMF}$. A cartoon of a cloud whose SFZ undergoes fragmentation and collapse is shown in Figure 5.

In the fragmented SFZ, the arrangement of MISs can be visualized as closely spaced "chains of cores" as observed in B213 (Tafalla & Hacar 2015) and Aquila (Konyves et al. 2015). The uniform inter-MIS gas density $n_{min}$ can be considered an average over the filament and inter-filament gas surrounding cores. In the limit of an infinitely extended isothermal medium, this inter-MIS gas has the pressure needed to keep each MIS critically stable.

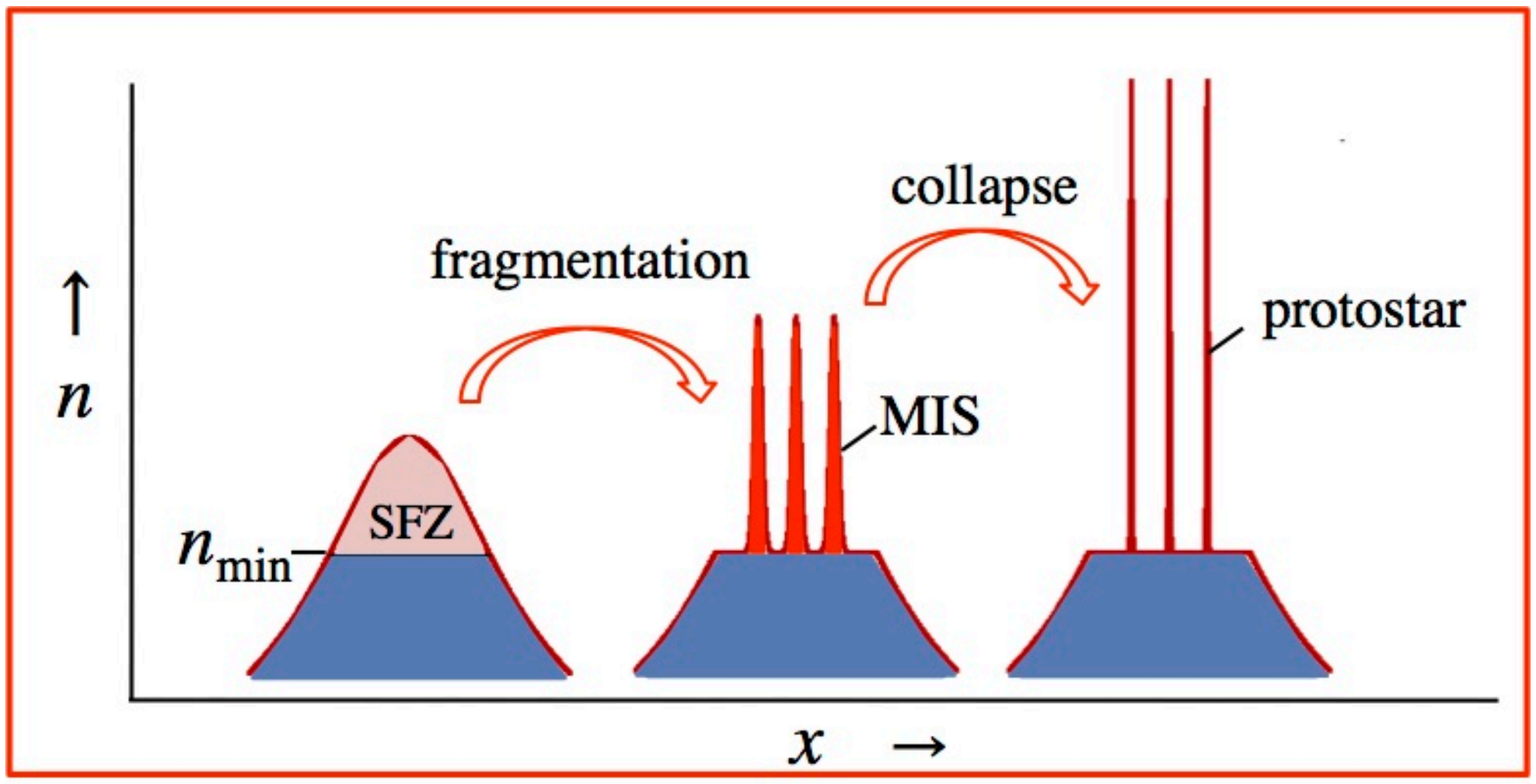

**Figure 5.** Cartoon of fragmentation and collapse in a star-forming zone (SFZ) having density $n \geq n_{\min}$. The SFZ in the initial cloud (*left*) fragments into MISs **in** a uniform medium of density $n_{\min}$ (*center*). Each MIS is a Bonnor-Ebert sphere which collapses to form a protostar whose final mass is the mean mass of the IMF (*right*).

The number of MISs in the fragmented SFZ, $N_S$, is found by mass and volume conservation from the unfragmented SFZ to the fragmented SFZ. The SFZ mass is $M_Z = \rho_{\min}(V_Z - N_S V_S) + N_S M_S$, or equivalently

$$N_S = \frac{V_Z}{V_S} f \quad , \qquad (27)$$

where $f$ is the volume filling factor of MISs,

$$f \equiv \frac{q_Z - 1}{q_S - 1} \qquad (28)$$

and where $0 \le f \le 1$ or $1 \le q_Z \le q_S = 2.46$. Thus $N_S$ depends only on the SFZ volume and its initial concentration $q_Z$, since the parameters $q_S$ and $V_S$ are constant properties of the MIS. $N_S$ can be understood as the number of MISs needed to make the mean density of the fragmented SFZ equal the mean density of the unfragmented SFZ.

The mean fragment spacing and star formation efficiency of the SFZ can be expressed solely in terms of the initial SFZ concentration $q_Z$ and on constant parameters. The mean fragment spacing in 3D is $\lambda_S = (V_Z / N_S)^{1/3}$, whence equations (27)-(28) give the spacing in terms of MIS radius as

$$\frac{\lambda_S}{R_S} = \left(\frac{4\pi}{3f}\right)^{1/3} , \qquad (29)$$

and in terms of the Jeans length for the mean SFZ density as

$$\frac{\lambda_S}{\lambda_J} = C_S \left(\frac{q_Z}{\pi}\right)^{1/2} \left(\frac{4\pi}{3f}\right)^{1/3} , \qquad (30)$$

where $C_S = R_S (G\rho_{min})^{1/2}/\sigma = 0.486$ is a property of the BE sphere with velocity dispersion $\sigma$ (McKee & Ostriker 2007 [MO07]).

The star formation efficiency, or ratio of the final mass in protostars to the initial SFZ mass, is given by

$$SFE = \varepsilon_S \frac{1 - q_Z^{-1}}{1 - q_S^{-1}} \quad . \tag{31}$$

This SFE refers only to new stars which can form in the SFZ, and does not include already formed stars in the SFZ. It is meaningful only when a substantial number of MISs are predicted to form. The relations in equations (29) - (31) are shown in Figure 6.

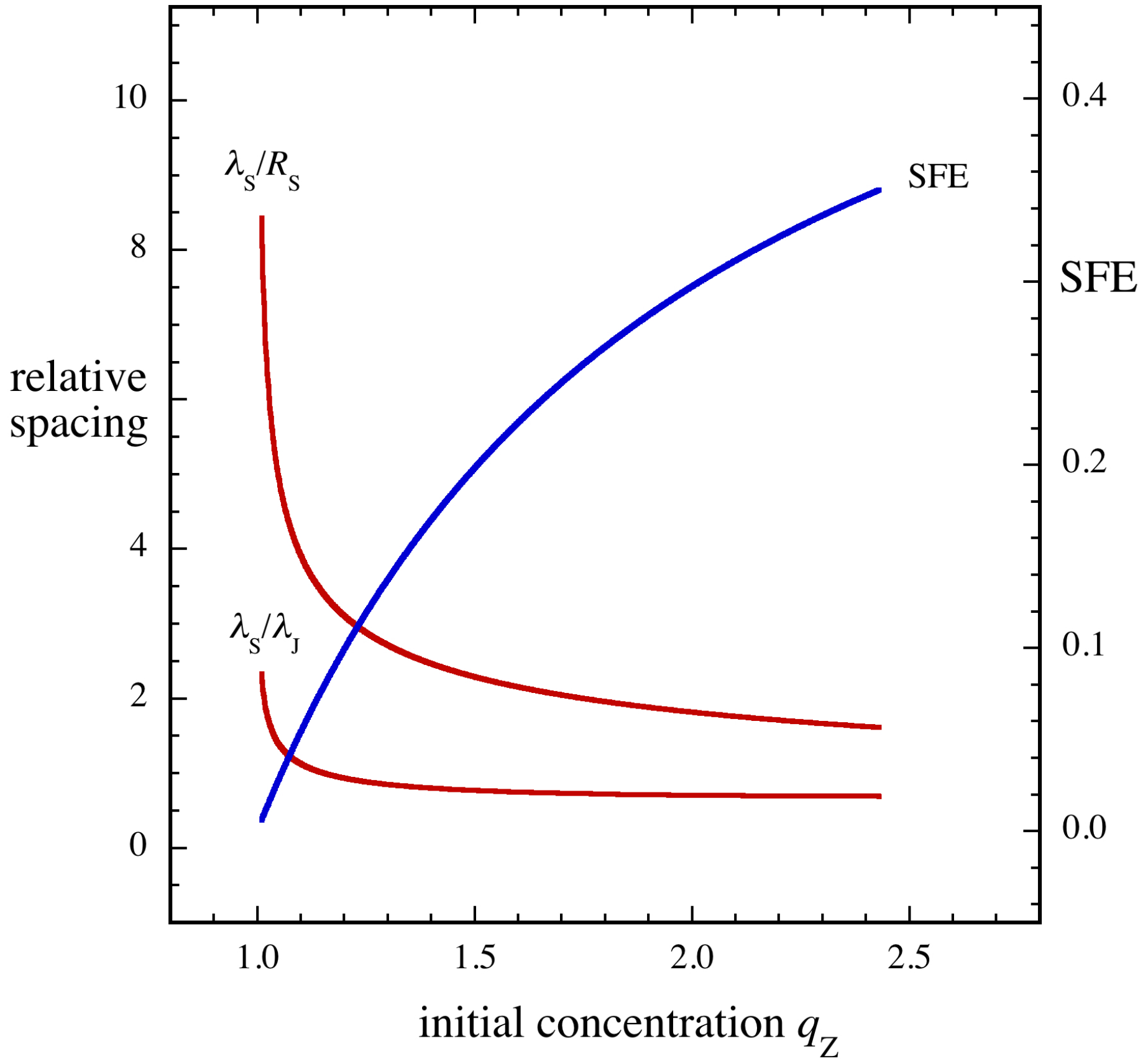

**Figure 6.** Mean fragment spacing in terms of radius ($\lambda/R_S$) and initial Jeans length ($\lambda/\lambda_J$) (*red curves*) and star formation efficiency (SFE, *blue curve*), for a fragmented star formation zone as functions of its initial concentration $q_Z \equiv \bar{n}/n_{min}$.

Figure 6 and equations (29) - (31) show that MISs in the fragmented SFZ have typical spacing about three MIS radii, or about one Jeans length for the mean density of the initial SFZ. The predicted SFE ranges from 0 to 0.3. These SFE values span the range of observed estimates for five large clouds with extinction $A_V >\sim 2$ mag, where the SFE is 3-6 % (Evans et al. 2009), to smaller zones within these same clouds with $A_V >\sim 6$ mag, where the SFE is 5-15 % (Jørgensen et al. 2008), to deeply embedded clusters, where the SFE is 10-30% (Lada & Lada 2003, G09). Analysis of simulations and observations indicates that the SFE generally increases with decreasing size scale until one reaches the dense core scale (Federrath & Klessen 2013; Padoan et al. 2014), in accord with this fragmentation model.

3.3. Comparison of SFZ fragmentation model with cluster observations

This SFZ fragmentation model applies only to regions whose concentration is less than the concentration of a single MIS, $q_{Z,max} = q_S = 2.46$ (MO07). This limit is probably consistent with most star-forming regions. The typical range of $q_Z$ in embedded clusters lies below this limit, according to analysis of a mid-infrared survey of embedded clusters within 1 kpc (Gutermuth et al. 2009, hereafter G09). For 27 of these clusters, the concentration $q_Z$ was obtained from the ratio of the peak to mean extinction, assuming that the gas density in each cluster follows a truncated Plummer-like sphere. The resulting concentration range is $1.00 \leq q_Z \leq 2.05$, which lies below the MIS limit $q_{Z,max} = 2.46$.

The relation between initial gas concentration and MIS spacing predicted in equation (29) is supported by the above cluster data, for embedded clusters in the sample of G09. For 22 values of $q_Z$ the predicted 3D spacing of MISs, $\lambda_S$, has mean ± standard error 0.14 pc ± 0.01 pc. The median projected spacing of the protostars in each cluster given in Table 8 of G09 is denoted here as $\lambda_{P2}$. This 2D spacing was converted to a 3D spacing $\lambda_{P3}$ by assuming that the effective cluster radius $R_{hull}$ in G09 Table 8 encloses a spherically symmetric distribution of stars, i.e. $\lambda_{P3} = \lambda_{P2}[4R_{hull}/(3\lambda_{P2}]^{1/3}$. The mean ± standard error of these 22 $\lambda_{P3}$ values is once again 0.14 ± 0.01 pc, in good agreement with the typical predicted MIS spacing.

For the same cluster sample the typical predicted and observed values of SFE also agree within statistical uncertainty. The predicted SFE has mean ± standard error 0.14 ± 0.01, based on equation (29) and on the 22 values of $q_Z$ from G09. The observed SFE has mean ± standard error 0.16 ± 0.02 based on 22 values of mean $A_K$ and mean surface density of protostars from G09.

This young cluster sample shows substantial consistency between the mean predicted spacing of MISs and the observed spacing of protostars, and between the mean predicted and observed SFE. The interpretation of this consistency depends on the birth time distributions of the observed and predicted protostars. If future protostars have a birth history similar to that of the already formed stars, the consistency suggests that the model accurately describes a steady-state cluster, where the rates of gas mass gain and depletion in the SFZ are equal, possibly due to accretion, star formation and feedback (Fletcher & Stahler 1994, Myers 2014). However if instead the model predicts only the number of protostars in one generation, while the observed population is the sum of many generations, the consistency may indicate that the model predicts too many stars for one generation.

The agreements between mean values of spacings, and of SFEs, do not extend to correlation between the individual predicted and observed spacings, or to correlation between the individual predicted and observed SFEs, from one cluster to the next. This lack of correlation may occur because the present inference of cluster gas concentration and of protostar spacing relies on the assumption of spherical symmetric distributions of gas and protostars. For young clusters this assumption is substantially more accurate on average than in individual cases. A more accurate analysis would require a more accurate model of the shape of each SFZ, as is done in the following individual cloud models.

## 4. Application to observed filamentary clouds

In this section each of the three models introduced in Section 2 is applied to an observed filamentary cloud of similar shape. The choice of model is dictated by the prominence of the filament's central bulge, increasing from TPS to STPS to TPPC. The observed cloud shapes presented here are sufficiently distinct that the choice of the model with the best-matching contours can be made visually for each cloud. In closer cases, it may be preferable to choose the best model by least-squares fitting of model and observed column density maps. In that case, it will be important to select the model whose fit parameter values have the least uncertainty.

### 4.1. TPS model of the Musca filament

The Musca filament appears as one of the simplest filaments among nearby clouds, spanning some 6 pc at a distance of ~150 pc (Knude & Hog 1998). Its central ridge and its embedded cores have column densities typical of nearby star-forming regions, yet it has remarkably little star formation, with one T Tauri star candidate (Vilas-Boas et al. 1994). According to detailed near-infrared observations of its extinction structure, its central region resembles an elongated ridge with a modest axial concentration (Kainulainen et al. 2015).

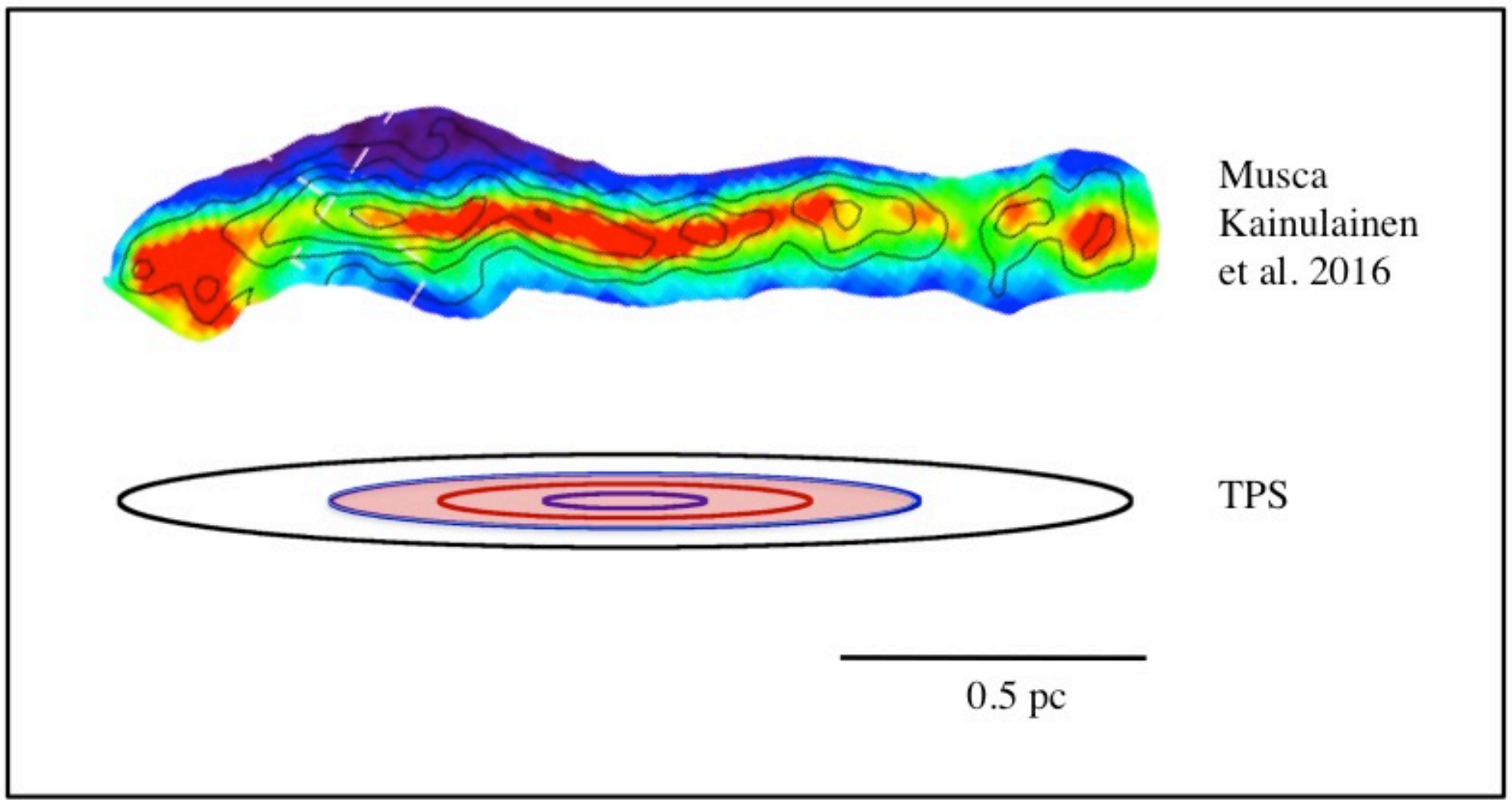

**Figure 7.** Observed column density contours of the central region of the Musca filament (Kainulainen et al. 2016) and TPS model contours, increasing from 4.5 $10^{21}$ $cm^{-2}$ in steps of 3 $10^{21}$ $cm^{-2}$. Shading indicates the projected "star formation zone" whose gas is considered dense enough to form low-mass stars.

The TPS model parameters are determined by identifying the lowest well-defined contour level as the background column density $N_{bk}$, and by measuring its projected length $L$ and width $2R$. The filament axis is assumed to lie in the plane of the sky. The HWHM $x_{1/2}$ is measured from the radial $N$-profile through the position of peak column density $N_{max}$. Then model parameters are obtained from equation (18) and from $a = L/(2R)$. For the observed properties $N_{bk}$ = 4.5 $10^{21}$ $cm^{-2}$, $N_{max}$ = 19 $10^{21}$ $cm^{-2}$, $R$ = 0.075 pc, $L$ = 1.6 pc, and $x_{1/2}$ = 0.035 pc, the derived model properties are $r_0$ = 0.027 pc, $n_0$ = 7.1 $10^4$ $cm^{-3}$, $\Xi$ = 2.78, and $a$ = 10.9. The contour maps of the the observed Musca filament central region and its TPS model are shown in Figure 6.

The model contours in Figure 7 approximate the large-scale shape and value of the observed contours, within limitations due to the assumed axisymmetry. They reflect the observed filament width and elongation, and the increase in column density from the edge to the central ridge.

4.2. Star forming zone in the Musca TPS model

The boundary of the SFZ in a TPS model is a prolate spheroid defined by

$$[\xi^2 + \eta^2 + (\zeta / a)^2]_b = \Xi_Z^2 \tag{32}$$

where

$$\Xi_Z^2 \equiv \frac{n_0}{n_{Z,\min}} - 1 \tag{33}$$

in analogy with equations (5) and (17), and where $n_{Z,\min} = n_{S,\min}$. The corresponding column density contour is obtained from equations (17) and (18) where $\mu = \Xi_Z$.

For the Musca model parameters, $\Xi_Z = 2.11$, givings $N_Z = 8\ 10^{21}\ \text{cm}^{-2}$, as shown in Figure 6. The mass of gas denser than $n_{\min}$ is then $M_Z = 10.8\ M_{Sun}$. However, most of this mass is at the tapering ends of the filament model, where the radial extent is less than that of a MIS. Considering only gas whose radial extent exceeds $R_{MIS}$, the SFZ can harbor ~3 MISs. Thus the central zone of the Musca filament has enough mass and extent of dense gas to form at most a few low-mass stars. Since this zone has no known protostars, it **may** be considered in an early stage of its star formation history.

It is also possible that the central Musca filament is presently starless, and will remain starless, because it is magnetically subcritical. However, no measurements of magnetic field strength in this region are presently available. On the other hand, the high density and nearly thermal velocity dispersion of the Musca filament gas make it a good candidate for low-mass star formation (Kainulainen et al. 2014).

The mean spacing of MISs in this fragmentation model is 0.1 pc according to equation (29). This spacing is similar to the mean spacing 0.08 pc of the four dense cores in the B213 filament (Tafalla & Hacar 2015), and to the median spacing 0.09 pc of cores in Aquila (André et al 2014). The Musca central region was identified as being in the process of fragmentation by Kainulainen et al. (2016). Their analysis in terms of global instability in an isothermal filament implied fragment spacing ~ 0.4 pc. A similar analysis of the infrared dark cloud G11.11-0.12 gave spacing as 0.2 pc due to local Jeans instability and 0.4 pc due to global instability (Kainulainen et al. 2013).

4.3. STPS model of the L43 filament

The L43 filament in northern Ophiuchus is about 1.6 pc long, with a central dense core harboring the YSO RNO 91 and an associated CO outflow. A second YSO, RNO 91, is located a few 0.1 pc away (Lynds 1962, Mathieu et al. 1988, Benson & Myers 1989, Reipurth 2008, Lombardi et al. 2008). Figure 8 shows a contour map of its large-scale structure based on near-infrared extinction of background stars (Dobashi et al. 2011), and the corresponding STPS model.

This STPS model shape was chosen because L43 is corelike on small scales and filamentary on large scales, with greater axial concentration than the Musca filament in Figure 6. The model parameters were obtained with a procedure similar to that for the Musca filament. However the extinction map resolution is several arcmin, so the peak is poorly resolved and the peak column density $N_{max}$ is poorly determined. Instead the adopted model column density $N_{max}$ = 10 $10^{21}$ cm$^{-2}$ was inferred by adjusting $\Xi$ until the lower-density model contour positions approximately match those observed. This value of $N_{max}$ is a compromise between the peak value of the extinction map and the peak value of a higher-resolution observervation (Chen et al. 2009). For the observed properties $N_{bk}$ = 1 $10^{21}$ cm$^{-2}$, $N_{max}$ = 10 $10^{21}$ cm$^{-2}$, $R$ = 0.20 pc, and $L$ = 1.6 pc, the derived model properties are $r_0$ = 0.067 pc, $n_0$ = 1.7 $10^4$ cm$^{-3}$, $\Xi$ = 3.0, and $a_{max}$ = 4.0.

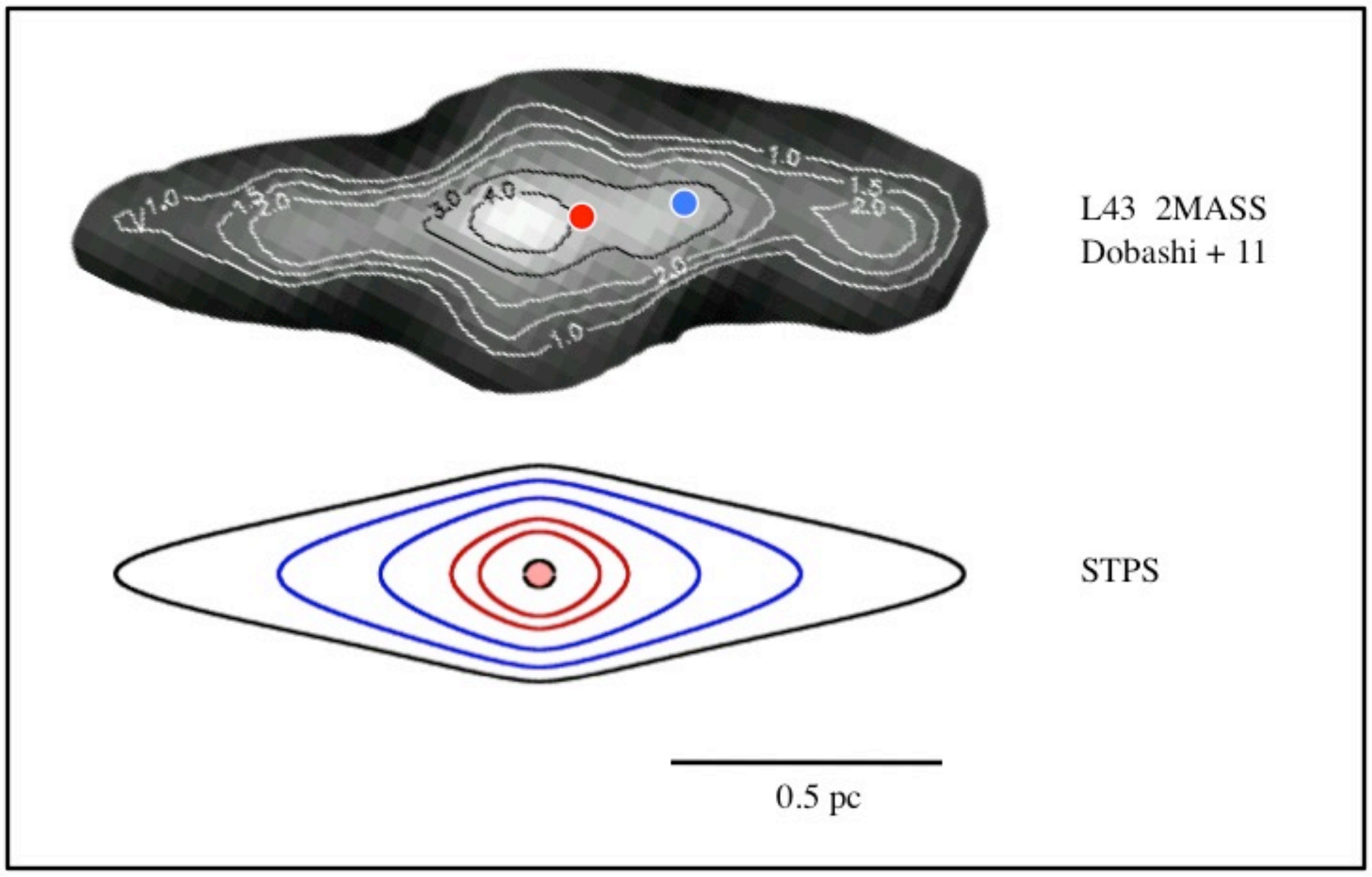

**Figure 8.** Contours of column density in L43, ranging from 1 $10^{21}$ cm$^{-2}$ in steps of 1 $10^{21}$ cm$^{-2}$, according to extinction of 2MASS sources (Dobashi et al. 2011), and according to the STPS model. The red circle indicates the protostar RNO91 and the blue circle indicates the YSO RNO90. The model zone of gas dense enough for further low-mass star formation is the small shaded circle, where the column density exceeds 8 $10^{21}$ cm$^{-2}$.

The SFZ in L43 was obtained as in Section 3.4. Its bounding column density is 8 $10^{21}$ cm$^{-2}$ as in Musca. However the extent of this SFZ, shown by the shading in Figure 7, is much smaller than in the Musca filament, and is smaller than the extent of a single MIS. Thus this SFZ does not have enough dense gas to make a single low-mass star, assuming the usual value of efficiency $\varepsilon_S$. The large values of $\lambda_S/R_S$ and $\lambda_S/\lambda_J$ in Figure 5 indicate that this SFZ is too small to host another fragment. Although finer resolution is desirable, these conclusions are not likely to change as a result of improved resolution. This result implies that L43, which has already formed two low-mass stars, is unlikely to form further low-mass stars. Consequently it appears to be near the end of its star-forming history, in contrast to the Musca filament.

4.4. TPPC model of the Coronet filament

The Coronet cluster is a dense group of eight protostars and five YSO candidates extended over ~ 0.1 pc in the R CrA complex, which harbors some 116 protostars and YSO candidates over ~ 3.2 pc. These population numbers are based primarily on *Spitzer* observations (Peterson et al. 2011). The dense gas in the complex is filamentary, with one main filament and several side branches. This region differs from the Musca and L43 filaments discussed above, since it has much greater peak column density, reaching ~ 45 $10^{21}$ $cm^{-2}$ (Chini et al. 2003, Alves et al. 2014).

The dense gas of the Coronet and its filamentary environment (Chini et al. 2003) harbors 10 protostars and 10 YSOS (Peterson et al. 2011). Its central gas temperature is ~ 20 K, decreasing to ~10 K according to $NH_3$ line observations (Kontinen et al. 2003), in contrast to the more nearly isothermal gas in Musca and L43. Its map appearance is dominated by the central core, with relatively faint filamentary extensions. This shape differs from both the TPS and STPS shapes, but matches more closely the TPPC model shape. The observed 1.2 mm emission map and associated young stars are shown in Figure 9, along with the TPPC model column density map.

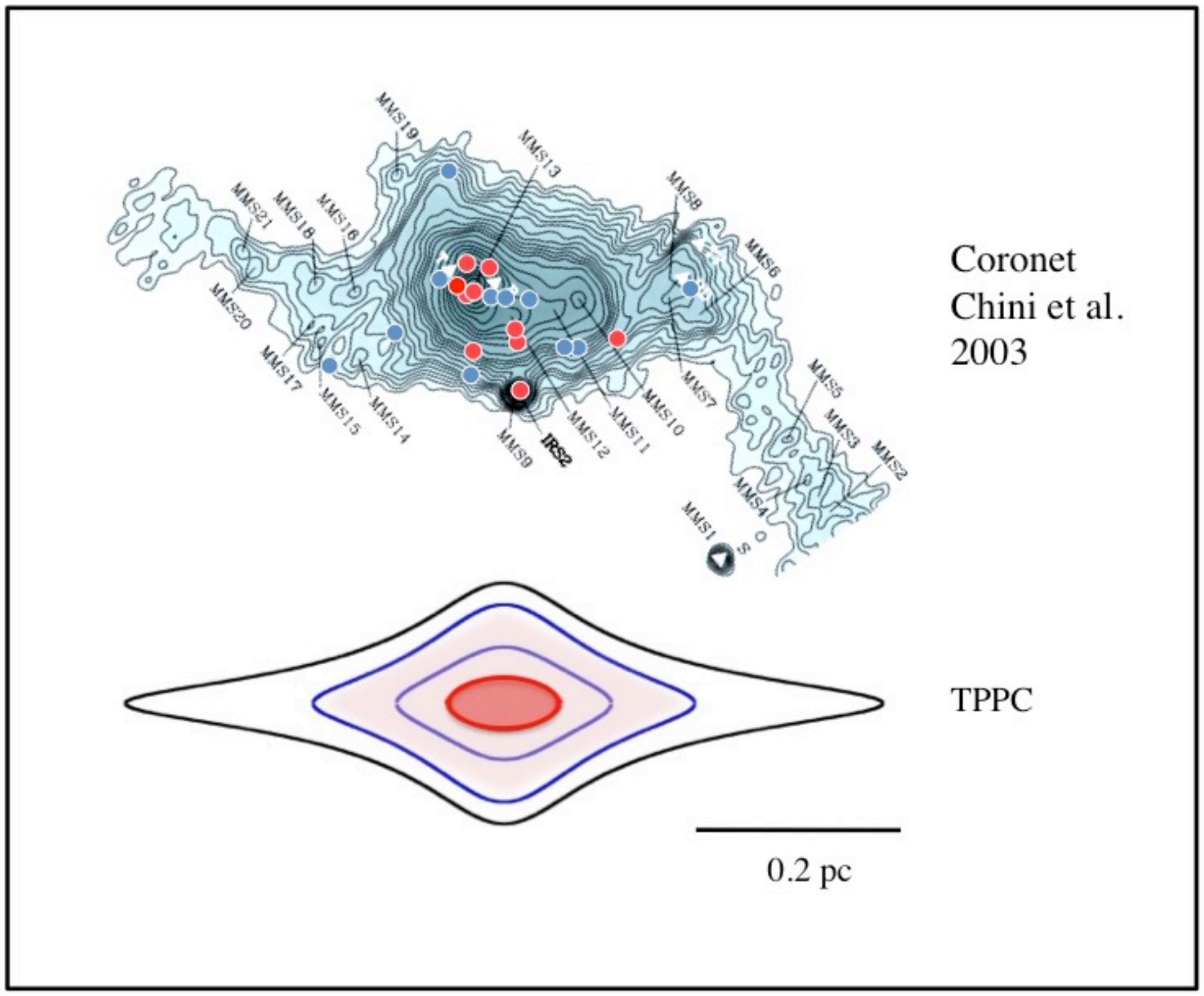

**Figure 9.** Maps of the Coronet cluster and associated filamentary cloud, at 1.2 mm wavelength (Chini et al. 2003), and according to a TPPC model. The red and blue symbols indicate all associated protostars (10 Class I and flat-spectrum) and YSOs (10 Class II) projected on the contour map (Peterson et al. 2011). The 1.2 mm contours extend from 0.10 to 2.8 Jy beam$^{-1}$ (Chini et al. 2003), corresponding to column density 1.6 to 45 $10^{21}$ cm$^{-3}$ (Alves et al. 2014). The TPPC column density contours represent 1.6, 6.1, 15, and 29 $10^{21}$ cm$^{-2}$. The light and dark shading in the TPPC map indicates the star forming zone as in Figures 6 and 7, for gas kinetic temperatures $T_S$ = 10 K and 20 K. Most of the observed protostars and YSOs lie in the model star formation zone.

The TPPC model parameters were obtained with the same procedure as for the TPS model of the Musca filament in Section 3.4. The lowest map contour was set as the bounding contour. The length $L$ was measured along the map ridge. The radius $R$ = 0.22 pc and half-maximum radius $x_{1/2}$ = 0.034 pc were measured along a line perpendicular to the long axis through the map peak. The maximum column density peak is 45 $10^{21}$ $cm^{-2}$, following the determination of the peak extinction as $A_K$ = 5.4 by Alves et al. (2014). These properties were used to obtain the model parameters as $a$ = 3.1, $\Xi$ = 4.85, $r_0$ = 0.024 pc, and $n_0$ = 2.2 $10^5$ $cm^{-3}$. The contours of the axisymmetric model depart from the more complex observed shape, but the model contours capture the dominant central concentration and faint filamentary extensions of the observed core-filament system.

The model star formation zone is presented in Figure 9 for gas temperatures spanning the values derived from $NH_3$ observations, from 10 K and $n_{Z,min}$ = 1.3 $10^4$ $cm^{-3}$ (light shading) to 20 K with $n_{Z,min}$ = 1.0 $10^5$ $cm^{-3}$ (dark shading). The 20 K part of the SFZ corresponds roughly to the Coronet cluster with closer protostar spacings, but it only contains enough dense gas mass to make one new low-mass star according to equations (27)-(28). The 10 K part corresponds to the surrounding, more filamentary region with more evolved YSOs having greater spacings. It contains enough dense gas mass to make ~7 new low-mass stars, with mean spacing 0.09 pc and SFE = 0.3.

These models of SFZ fragmentation suggest that the Coronet region may be near the middle of its star forming history, since it has enough dense gas to add ~8 protostars to the 10 protostars and 10 YSOs now known. The Coronet contrasts with the Musca central region, which appears near the start of its star-forming life, since it has no protostars or YSOs, but has enough dense gas to add ~ 3 protostars. The Coronet also contrasts with L43, which appears near the end of its star-forming life, since it has formed one protostar and one YSO, but has too little dense gas to form any further stars.

## 5. Summary and discussion

### 5.1. Summary

The main points of this paper are

(1) Three axisymmetric models of core-filament density structure are presented to describe large-scale filament properties, and to improve our understanding of star formation.

(2) These models are more realistic than 1D Plummer cylinder (PC) models often used to interpret filament observations. They can match the finite length and mass of observed filaments, and they can include embedded cores. They resemble observed column density contour maps more closely than do PC contour maps, and at high $N$ their $N$-pdfs are pole-free power laws, like observed $N$-pdfs but unlike PC $N$-pdfs.

(3) Each model allows identification of a "star-forming zone" (SFZ) whose mean density matches that of star-forming dense cores and whose column density exceeds the "star formation threshold" ~ 6 $10^{21}$ $cm^{-2}$. This zone is modelled as gas denser than $n_{min}$, the minimum density of a "MIS," a 1 $M_{Sun}$ BE sphere which forms a star of mean IMF mass $\overline{M}_{IMF} = 0.36\ M_{Sun}$.

(4) The number $N_S$ of new low-mass stars which can form in a SFZ is predicted by assuming that the initial SFZ fragments into $N_S$ MISs in a uniform medium of density $n_{min}$. In this thermal fragmentation model, the stars which can form have mean spacing and star formation efficiency depending only on the concentration $q_Z \equiv \bar{n} / n_{min}$ of the initial SFZ.

(5) The gas concentrations, protostar spacings, and star formation efficiency (SFE) in a sample of 22 embedded clusters (G09) match properties of the SFZ fragmentation model. The range of $q_Z$, $1.00 \leq q_Z \leq 2.05$, lies within the range of allowed values 1 to 2.43. The typical protostar spacing, 0.14 pc, and the typical star formation efficiency, 0.16, each agree within statistical error with the typical predicted value. The typical spacing is close to the Jeans length for the mean density of the initial SFZ, as expected for a fragmentation model based on thermal gas properties.

(6) Application to filamentary clouds in L43, Musca Center, and the Coronet provide simple models of their large-scale structure of density and column density. In turn, the models of the SFZ and its fragmentation indicate that the Musca Center filament is dense enough to form its first few low-mass stars, the Coronet can add some ~8 stars to the ~20 already known, but L43 has too little dense gas to add any new stars to the two already known. These results suggest that Musca Central is near the start of its star-forming life, the Coronet is near the middle, and L43 is near the end.

5.2 Limitations

The models presented in this paper should be applied with understanding of their limitations.

The TPCC, TPS, and STPS density models are axisymmetric and centrally condensed, so they cannot describe nonaxisymmetric structure or filament fibers or networks. At best these models are large-scale averages over filament position and velocity.

The mean radial $N$-profiles of the TPCC, TPS and STPS models approximate the shape of $p = 2$ Plummer-like cylinder profiles used to fit observed $N$-profiles (A11, A15), but they depart significantly when the region included in the average extends to positions whose peak column density falls below 0.1-0.2 of the central column density.

The three models presented here are variants of Plummer-like structures only with $p = 2$. The comparison of their $N$-profiles with PC profiles remains untested for $p$ values in the broader range ~ 1.3 - 2.4 inferred from observations (A11). Similarly, no comparison of high-$N$ power-law slopes was made between these three models and other models with $p \neq 2$, or between these three models and $N$-pdfs of observed filamentary regions.

The properties of the SFZ are derived assuming that the mass of the MIS is 1 $M_{\rm Sun}$, based on core-star efficiency estimates $\varepsilon_{\rm S}$ = 0.3-0.4 which follow from comparing the core mass function and the IMF (Alves et al. 2007, Könyves et al. 2015). However estimates of $\varepsilon_{\rm S}$ based on counting protostars in Ophiuchus and Perseus give a lower value, $\varepsilon_S$ = 0.13-0.17 (Jørgensen et al. 2008). Adopting this value would decrease $n_{min}$ by a factor ~4, reducing the applicability of the SFZ and fragmentation models.

The SFZ fragmentation model is restricted to BE spheres which make protostars of a typical IMF mass, so it cannot account for formation of massive stars, and for the role of massive stars in heating and dispersing the star-forming gas.

The SFZ fragmentation model does not explain the process which transforms the unfragmented SFZ into the fragmented SFZ. One suggestion requires supersonic anisotropic converging flows along magnetic field lines, which form filaments having core seeds at birth. The cores grow in ≤ ~1 Myr (Chen & Ostriker 2015). A purely thermal mechanism is "geometrical fragmentation" where a straight isothermal filament is subject to small-amplitude sinusoidal bending. For central density > 5 $10^4$ $cm^{-3}$, the filament forms self-gravitating cores at the bends in ~ 1 Myr (Gritschneder et al. 2016). It may be useful to test the SFZ fragmentation model with numerical simulations, starting from the cloud models presented here as initial states.

### 5.3. Thermal fragmentation

It seems surprising that the simple model of thermal fragmentation in Section 4 predicts typical cluster spacing and SFE values which match those derived from observations of young clusters by G09. In low-density gas, with 2 < log $n$ < 4, supersonic turbulent motions are believed essential to form filaments and to prevent stars from forming too rapidly (Vazquez-Semadeni 1994, Klessen & Burkert 2001, Federrath 2015). However in denser star-forming gas with log $n$ > 4, thermal physics appears sufficient to describe many features of star formation, perhaps because turbulent motions and magnetic forces have become less important than thermal pressure and gravity (Larson 2005).

This view is supported by an increasing number of observations which reveal that some dense regions with "turbulent" line widths at low resolution have sonic or transsonic line widths at high resolution (Pineda et al. 2010, Hacar et al. 2013, Kainulainen et al. 2015). It is also supported by a study of dense cores in regions of massive star formation, indicating that at high resolution their number of fragments correlates better with their number of thermal Jeans masses than with their number of turbulent Jeans masses (Palau et al. 2015).

On the other hand, the number and masses of stars expected to form in a SFZ must also depend on the initial spatial distribution of the SFZ gas. A spherical SFZ has a deeper potential well than an elongated SFZ having the same mass and the same number of thermal Jeans masses. The deeper well will collapse to a single massive object, while the shallower well will tend to first produce less massive objects at its ends, which later fall toward each other (Nelson & Papaloizou 1993).

Thus in the thermal fragmentation model presented here, formation of the predicted number of objects seems to require a decentralized gas distribution dominated by many small wells. Then nearly all star-forming collapses are "local" and only a small fraction gain significant mass from "global" collapse (Wang et al. 2010). One may speculate that a SFZ whose protostars have mean spacing matching the thermal Jeans length must arise from flows and gravity which structure most of the SFZ into filaments and BE-like cores. The ordered and chaotic structure of these flows is therefore of great interest. Once the SFZ has this internal structure, it can produce the number of low-mass stars predicted by a simple thermal fragmentation picture.

The similarity of the model fragment spacings and the thermal Jeans length noted in Section 3 reflects the thermal physics assumptions of this fragmentation model. It does not imply that the fragments form by a thermal Jeans instability. Instead the similarity of lengths is due to the definition of the initial SFZ as gas which is denser than $n_{\rm min}$, the minimum density of a BE sphere. Since the Jeans length and the BE sphere diameter have the same dependence on temperature and mean density, with slightly different coefficients, these SFZ definitions guarantee that fragment spacing will approximate the Jeans length for the mean SFZ density.

In Section 3, the Jeans length used for comparison with predicted MIS spacings is the "local" Jeans length for fragmentation of a uniform medium with thermal velocity disperson $\sigma$ and density $\rho$, $\lambda_J = \sigma[\pi/(G\rho)^{1/2}]$, and not the "global" Jeans length for fragmentation of an infinite, self-gravitating isothermal cylinder, $\lambda_{cyl} = 1.25\ 2^{3/2}\sigma[\pi/(G\rho_0)^{1/2}]$, where $\rho_0$ is its maximum density (Larson 1985, MO07, Kainulainen et al. 2013). At the temperature and density values considered here, the local value is ~0.1 pc and is closer to observed core spacings than the global value, which is significantly greater (André et al. 2014, Kainulainen et al. 2013).

## 5.4. Model applications

The models presented here describe the large-scale density structure of observed star-forming clouds more realistically than is possible with 1D models such as the Plummer cylinder, as discussed in Section 2. This improved description of cloud density structure allows identification of the "star forming zone " (SFZ) as gas denser than the minimum density of the Bonnor-Ebert sphere which makes a star of mean IMF mass, as discussed in Section 3.

This definition ties observed properties of star-forming gas - dense core volume density and threshold column density - to the typical IMF mass, implementing the ideas of Larson (2005) and Bate & Bonnell (2005). Identification of the SFZ allows a fragmentation model to estimate how many low-mass stars can form from the available dense gas. This estimate can discriminate star-forming clouds which are in the early, middle, or late stages of their star-forming history, as illustrated in Section 4.

A second application of these models is to provide more realistic initial conditions for tests of fragmentation models against simulations. Many simulations of star-forming fragmentation rely on turbulent driving or colliding flows to generate fragmentation, but their products do not necessarily resemble the filamentary clouds analyzed here. Other simulations start with simple geometric structures which lack the central concentration of observed clouds. The cloud models described here can extend the range of initial conditions for simulations beyond those considered by Nelson & Papaloizou (1993), Sigalotti & Klapp (2001), and by Burkert & Hartmann (2004).

## 6. Acknowledgements

Discussions with João Alves, Philippe André, Shantanu Basu, Andi Burkert, Blakesley Burkhart, Alyssa Goodman, Jouni Kainulainen, Vera Könyves, Charlie Lada, Nicola Schneider, Zach Slepian, Jürgen Steinaker, and Qizhou Zhang are gratefully acknowledged. Jouni Kainulainen and Nicola Schneider also provided useful *N*-pdf data. Continuing support from Terry Marshall is gratefully acknowledged. The referee made useful comments and suggestions which improved the paper.